%Paper: hep-th/9303143
%From: samson@guinness.ias.edu (Samson Shatashvili)
%Date: Thu, 25 Mar 93 13:48:10 EST

\input harvmac.tex

\def\frac#1#2{{#1\over#2}}

\def\exp{{\rm exp}}

\def\ch{{\rm cosh}}

\mathchardef\ka="101A

\catcode`\@=11
\def\slash#1{\mathord{\mathpalette\c@ncel{#1}}}
\overfullrule=0pt

\def\steepslash{\c@ncel}
\def\frac#1#2{{#1\over #2}}

\def\inbar{\,\vrule height1.5ex width.4pt depth0pt}
\def\IB{\relax{\rm I\kern-.18em B}}
\def\IC{\relax\hbox{$\inbar\kern-.3em{\rm C}$}}
\def\IP{\relax{\rm I\kern-.18em P}}
\def\IR{\relax{\rm I\kern-.18em R}}
\def\IZ{\relax\ifmmode\mathchoice
{\hbox{Z\kern-.4em Z}}{\hbox{Z\kern-.4em Z}}
{\lower.9pt\hbox{Z\kern-.4em Z}}
{\lower1.2pt\hbox{Z\kern-.4em Z}}\else{Z\kern-.4em Z}\fi}

\def\brs{{\scriptscriptstyle\rm BRS}}

\def\dda{\partial_{\alpha}}
\def\ddb{\partial_{\beta}}
\catcode`\@=12
\def\tr{{\rm tr}}

\Title{\vbox{\baselineskip12pt\hbox{IASSNS-HEP-93/15}
                \hbox{}}} {\vbox{\centerline{Comment on the Background}
\vskip6pt\centerline{Independent Open String Theory}}}

\bigskip\centerline{Samson L. Shatashvili \footnote{$\dagger$}
{Research supported by DOE grant DE-FG02-90ER40542. }
\footnote{$^\#$}{On leave of absence from St. Petersburg
Branch of Mathematical Institute (LOMI), Fontanka 27, St.Petersburg
191011, Russia.}}
\bigskip\centerline{School of Natural Sciences}
\centerline{Institute for Advanced Study}
\centerline{Olden Lane}
\centerline{Princeton, NJ 08540}
\vskip .5in

A direct derivation of the string field theory action in the Witten's
background independent
open string theory for the case when ghosts and matter are decoupled is given
and consequences are discussed.

\Date{March, 92}

Recently, a background independent formulation of open string field theory
was given in \ref\w1{E. Witten, Phys.Rev., D46 (1992) 5446.}.
This theory is formally
defined in the space of all two dimensional
world sheet theories, with the
world sheet lagrangian given by

\eqn\defi{L=\int_{\Sigma}d^2z(\frac{1}{8\pi}g^{\alpha\beta}{\dda}X^i{\ddb}X^j
\eta_{ij}+\frac{1}{2\pi}b^{ij}D_ic_j)+\int_{\partial\Sigma}d{\theta}V(X,b,c)}
Here, the first term is the closed string background
and the second term describes
an arbitrary boundary interaction with the condition that
the boundary operator $V$ has the form\foot
{For definition of $b_{-1}$ see [1].}

\eqn\deff{V=b_{-1}O,}
where O is a general operator of the ghost number 1.
If we denote by $O_i$ some basis in the space of such operators,
 $O=\omega^iO_i$, the basic definition of the string field theory action S
can be given via an equation that involves the two-point function
of $O_i$'s on the world-sheet:

\eqn\defff{dS=\frac{1}{2}\int_0^{2\pi}d{\theta_1}d{\theta_2}<dO({\theta_1})
\{Q,O\}({\theta_2})>,}
where $dS=d{\omega^i}d/dt^iS$ and $dO=d{\omega^i}O_i$ and
$<...>$ denotes unnormalized correlation function.\foot{It can be shown, using
Ward identities, that the two point function
on the right hand side in \defff\ is a closed form, see [1].}

In this letter we will show that if the operator $V$ doesn't contain
ghosts and is a function of purely bosonic variables, the  right hand side
of \defff\ actually reduces to a one-point function $d<{\Omega}^iV_i>$
with $\Omega$ being a certain linear combination of couplings $\omega$.
{}From this statement it follows immediately that the
left hand side of \defff\
is indeed an exact form and that

\eqn\ans{S=({\Omega}^i\frac{d}{d{\omega}^i}+1)Z,}
where $Z$ is the world sheet partition function. We will give the expression
for $\Omega$ below (see (15), (16), (17)).

The fact that the action can be written in the form
\ans\ was demonstrated before in \ref\w2{E. Witten,
Some calculations in background independent off-shell string theory,
Princeton Preprint IASSNS-HEP-92/63.} by explicit calculations
for the case of quadratic interaction $V={\omega}X^2$; also in [2]
for general $V(X)$  the existence of a relation of type \ans\
was proven and the
algorithm for constructing the vector field in \ans\ was described.

Our approach differs from the one in [2]: we will not use the exactness
of the one form on the  right hand side of \defff\ but we will just reduce
the correlation function in \defff\ to a
total derivative of a one-point function.
Also, we will give an explicit formula for
the vector field in \ans\ and  it's relation to the
Virasoro generators on world sheet.
As a consequence of the definition \defff\ we will see
that the zeros of the vector field
$\Omega$, which define the string filed theory equations of motion (see [2])
\foot{The fact that the string field theory action
on the classical equations of motion is given by the
world-sheet bosonic partition function
previously was suspected in
\ref\frts{E. Fradkin and A. Tseytlin, Phys.Lett, B163 (1985) 123.}
\ref\ch{A. Abouelsaood, C. G. Callan, C. R. Nappi and S. A. Yost,
Nucl. Phys. B280 (1987) 599.} and  elaborated in terms of
theory under the discussion in [2].},
are linear functions of couplings; therefore
the equations of motion in the present formalizm
are also linear. At the end we will make some comments.
The alternative  derivation of the action functional
presented here and the above observation might be helpful for a more detailed
understanding of open string field theory.

To evaluate the correlation function in \defff\ we will write the BRST
operator in terms of the matter and ghost stress tensors $T_m, T_g$:

\eqn\brs{Q=\int_0^{2\pi}d{\theta}c(\theta)[T_m(\theta)+T_g(\theta)].}
Also from the assumption that ghosts and matter are
decoupled in the theory we
can write $O=cV(X)$ and thus the correlation
function is given by two terms

\eqn\basic{\eqalign{\frac{1}{2}\int_0^{2\pi}d{\theta_1}&
d{\theta_3}<c(\theta_1)dV(\theta_1)
c(\theta_3)\partial_{\theta_3}c(\theta_3)V({\theta_3})>\cr
&+\frac{1}{2}\int_0^{2\pi}d{\theta_1}d{\theta_2}d{\theta_3}<c(\theta_1)
dV(\theta_1)c(\theta_2)c(\theta_3)
[T_m(\theta_2),V(\theta_3)]>.}}
To proceed further we need to evaluate the ghost correlation
functions in \basic. For the general 3-point function we have:
\eqn\gh{<c(\theta_1)c(\theta_2)c(\theta_3)>=2(\sin(\theta_1-
\theta_2)-\sin(\theta_1-\theta_3)+
\sin(\theta_2-\theta_3)).}
This leads to a simple expression for the first correlator in \basic

\eqn\bass{\frac{1}{2}\int_0^{2\pi}<d{\theta_1}d{\theta_2}
(2\cos(\theta_1-\theta_2)-2)
<dV(\theta_1)V(\theta_2)>,}
and to three terms for the second correlator in \basic:

\eqn\basics{\eqalign{\frac{1}{2}\int_o^{2\pi}d{\theta_1}d{\theta_2}
[i(-&<e^{i\theta_1}dV(\theta_1)[L_{-1},V(\theta_2)]>-c.c.)+\cr
&+i(-<dV(\theta_1)[L_1,e^{-i\theta_2}V(\theta_2)]>-c.c)-\cr-
&2\sin(\theta_1-\theta_2)<dV(\theta_1)[L_0,V(\theta_2)]>].}}
Here we use the notation $L_n=\int_0^{2\pi}
d{\theta}e^{in\theta}T_m(\theta)$.

The last term in \basics\ can be simplified if we use the fact that
$[L_0,V(\theta)]={\partial}V(\theta)$
and thus after integration by parts over $\theta_2$ it leads to

\eqn\com{-\int_0^{2\pi}d{\theta_1}d{\theta_2}\cos(\theta_1-\theta_2)
<dV(\theta_1)V(\theta_2)>.}
At the same time we can also simplify the first term in
\basics\ just by using the observation
that the commutator term here is nothing but the
commutator of the Virasoro generator $L_{-1}$ with
the world sheet action (note that the bulk lagrangian
commutes with all Virasoro generators).
Thus in the path integral  for this correlation function we
could integrate by parts and because there is
no anomaly in the measure (the construction
of a background independent string field theory
assumes that we have a critical string) we get:

\eqn\first{\frac{i}{2}(<[L_{-1},\int_0^{2\pi}e^{i\theta}dV(\theta)]>-c.c.).}

Now we can combine \bass, \basics, \com\ and \first\ and due to some
nice cancellations and the identity

\eqn\simp{<\int_0^{2\pi}d{\theta_1}dV(\theta_1)\int_0^{2\pi}
d{\theta_2}V(\theta_2)>=
d<\int_0^{2\pi}d{\theta}V(\theta)>-dZ,}
we obtain

\eqn\good{\eqalign{dZ-&d<\int_0^{2\pi}d{\theta}V(\theta)>+\frac{i}{2}
(<[L_{-1},\int_0^{2\pi}d{\theta}e^{i\theta}
dV(\theta))]>\cr
&-<\int_0^{2\pi}d{\theta_1}dV(\theta_1)[L_1,\int_0^{2\pi}
d{\theta_2}e^{-i\theta_2}V(\theta_2)]>-
c.c),}}
where $Z$ denotes the world sheet partition function.

At this point we have to make an important assumption that
the exterior derivative in the space of couplings, d, commutes with Virasoro
generators $L_1, L_{-1}; [d,L]=0$. This property was crucial also in the
consideration of [1] (e.g. for the prove that the
symplectic form is closed and BRST invariant, see [1]).
It is important to keep in mind that,
because we are dealing in principle with non-renormalizable theory and
are using a
cut-off, we have to  require that short distance behavior
of the exact Green function, $G(z_1, z_2)$, in the
theory is still logarithmic;
this is certainly true on the bulk, but not
guaranteed when both $z_1$ and $z_2$
are on the boundary: $z_1=\exp(i\theta_1), z_2=\exp(i\theta_2)$.
The latter requirement leads to certain
restrictions on the
couplings $\omega^i$. Only in this case we can
define the Virasoro
algebra through the world sheet stress-tensor
both in the bulk and on the
boundary, and have
geometric transformation
properties for the boundary operators,
corresponding to reparametrization of the
world-sheet. Unfortunately, because generically we
will have unrenormalizable boundary interactions, we can not make this point
more elaborate. Although, we can make some comments if boundary
interaction is quadratic:
\foot{ I would like to thank E. Witten for pointing
out this possibility.}
$V=\sum_0^{\infty}\omega^{ij}_nV^{ij}_n+a;
V^{ij}_n=\int_0^{2\pi}d{\theta}X^i\partial^nX^j$,
and $a$ is a constant.
In this case, Green function with both points
on boundary is given by $<X^i(\theta_1)X^j
(\theta_2)>=
\sum_{k=-\infty}^{\infty}(|k|+\sum_{n=0}^{\infty}{\omega}_n(ik)^n)^{-1}_{ij}
\exp{i(\theta_1-\theta_2)}$.
Now if the infinite sum in the denominator for this Green function
goes to a constant
for large $k$, we will have a logarithmic short distance divergence and
we can define all operators $V$ by point splitting and subtracting
the divergent part, which will not depend on the couplings.\foot{The
treatment of regularization problems in the case when there are finite number
of non-zero couplings ${\omega}_n$ in
quadratic potential is given in \ref\keke{K. Li and E. Witten,
Princeton preprint IASSNS-HEP-93/7.}}

After the above comment we can now exchange the order of $d$ and
$L_{-1}(L_1)$ in
\good, and thus conclude that the right hand side of \defff\ is an exact form
and defines the string field theory action up to a constant:

\eqn\final{d(Z-<\int_0^{2\pi}d{\theta}V(\theta)>+\frac{i}{2}
(<\int_0^{2\pi}d{\theta}
e^{i\theta}[L_{-1},V(\theta)]>-c.c.))=dS.}
Thus, we have expressed the action S in terms of
a one-point function of the boundary
operator $V$. Let us represent $V$ as a sum of basic operators,
$V_i$, $V=\omega^iV_i$,
where $V_i$ corresponds to $O_i$ introduced above and is a function
of the field $X$ and it's derivatives. The operation $\delta$

\eqn\tr{{\delta}V_i=\frac{i}{2}[e^{i\theta}L_{-1}-e^{-i\theta}L_1, V_i]}
defines a linear transformation in the space
of operators $V_i$, or equivalently,
in the parameter space $\omega^i$:

\eqn\trans{\delta(\omega^iV_i)=\hat\omega^iV_i=\omega^i{\delta}V_i,}
and $\hat\omega$ is given by \tr\ and \trans.
This completes our derivation of desired formula for action S:

\eqn\finall{\eqalign{S=&(\Omega^i\frac{d}{d\omega^i}+1)Z;\cr
&\Omega^i=(\hat\omega^i-\omega^i),}}
and zeros of $\Omega$
\eqn\zero{\Omega^i=0}
are classical equations of motion.

After the equations of motions \zero\ are derived for general boundary
interactions (with restrictions on short distance behavior emphasized above)
we would like to make some comments. As it follows from \zero, \tr\ and
\trans, these equations are linear and gauge symmetries
are given by zero modes of the operator
$\delta-1$. In the notation
$V(X)=T(X)+A_{\mu}{\partial}X^{\mu}+...$ from \tr\ and  \trans\
the equation $\Omega=0$ leads to linearized tachyon equation, Yang-Mills
 equation and so on.
The non linear equations are coming
out from our consideration only if we restrict the class of boundary
interactions even more (or, if short distance behavior depends on a coupling
constant, as in the case of having only first derivatives of $X$ in $V$:
$V=\omega^{ij}X^i{\partial}X^j$ ). It would be very interesting to find the
physical requirements that lead to natural
restrictions on boundary interactions.
One might think that the reason of linearity
of equations of motion is the assumption
for operators $O$ being the linear functions of coupling constants:
$O={\omega}^iO_i$. If we do not impose this condition, our final
answer is still given by a one point function of the operator $(\delta-1)V$:

\eqn\may{S=<(\delta-1)V>+Z.}

In the case where we have the subspace of all boundary
interactions with only renormalizable (but nonlinear) interactions,
our assumptions about the short distance behavior of the exact Green function
are completely derivable. This suggests that the question of linearity
is not related to renormalization problems but rather
to definition \defff.
In the lines  of the original idea of [1],
when the string field theory action is defined by a BV antibracket, what
probably has to be done in order to get nonlinear
classical dynamics for general boundary interaction,
is that the definition of the antibracket has to be modified.

Acknowledgements: I would like to thank E. Verlinde for
collaboration in the initial
stage of this work and  J. Distler, K. Li, V. Periwal and
E. Witten for helpful discussions.

\listrefs

\end